\newcommand{\be}[0]{\begin{equation}}
\newcommand{\ee}[0]{\end{equation}}
\newcommand{\bea}[0]{\begin{eqnarray}}
\newcommand{\eea}[0]{\end{eqnarray}}

\documentclass[12pt]{article}

\usepackage{graphicx}
\usepackage{psfrag}

\begin{document}
\large
\hfill\vbox{\hbox{DCPT/05/42}
            \hbox{IPPP/05/21}}
\nopagebreak

\vspace{2.0cm}
\begin{center}
\LARGE
{\bf Estimating the $I=3/2$ $K\pi$ interaction\\ in $D$ decay}
\vspace{0.8cm}

\large{L. Edera and M.R. Pennington} 

\vspace{0.5cm}

{\it Institute for Particle Physics Phenomenology,\\ University of Durham, 
Durham, DH1 3LE, U.K.}\\
     
\vspace{1.0cm}

\end{center}

\centerline{\bf Abstract}

\vspace{0.3cm}

\small

\noindent Heavy flavour decay to light hadrons is the key to understanding
many aspects of the Standard Model from CP violation to strong dynamics. It is often presumed in line with 
the simple quark spectator model of $D$ decay to $K\pi\pi$ that the $K\pi$ system  has only $I=1/2$. 
E791 have recently presented an analysis of their results on $D^+\to(K^-\pi^+)\pi^+$ using a generalised 
isobar picture of two body interactions. While higher $K\pi$ waves are described by sums of known resonances, 
the $S$-wave amplitude and phase are determined bin-by-bin in $K\pi$ mass. The phase variation is found not 
to be that of $K^-\pi^+$ elastic scattering. This hints at a different mixture of $I=1/2$ and $I=3/2$ $S$-wave 
interactions than in elastic scattering.  Applying Watson's theorem to this generalised isobar model allows us 
to estimate the $I=3/2$ $K\pi$ $S$-wave component. We indeed find that this is larger than in  
hadronic scattering or semileptonic processes.

\normalsize

\vspace{0.5cm}
\newpage

\parskip=2mm
\baselineskip=5.6mm

\noindent Drawing quark line diagrams for heavy flavour decays provides a guide to the weak interaction dynamics that takes place. Thus as shown in Fig.~1a, in $D$ decay to $K\pi\pi$, the $c$ quark is seen to change into an $s$ quark by emitting an off-shell $W^+$ that materialises as a $\pi^+$. The other quark in the $D$, for instance a $d$, acts merely as a spectator and with the $s$ quark forms on the creation of a $u{\overline u}$ or $d{\overline d}$ a $K\pi$ system. Since the isospin of the $c$ and $s$ quarks is zero, it is then natural to assume that the $K\pi$ system has only the isospin of the $u$ or $d$ quark, namely $I=1/2$. This presumption is in keeping with a simple isobar picture of this decay in which well-known $I=1/2$ $K^*$ resonances, like the $K^*(890)$ or $K_2^*(1430)$, are readily seen. However, there are, of course, $\pi\pi$ resonances too, like the $\rho$, in other charge channels, that clearly indicate that the spectator picture is too simplistic. The $W$ can be internally converted into hadrons as shown in Fig.~1b, and then the $K\pi$ system can have $I=3/2$ as well as $I=1/2$. From wholly hadronic reactions we know the $I=3/2$ interactions do not produce resonances, in keeping with the $q{\overline q}$ model of hadrons. Moreover, in low energy hadron scattering unitarity is a key constraint and restricts the magnitude of every partial wave amplitude to be bounded by unity. However, in the weak decay reaction we have no such constraint. Whilst in $K$ decays we know there is the well-known $\Delta I=1/2$ rule, no such property is known for charmed decays. Consequently, we cannot presume that the $I=1/2$ component always dominate over $I=3/2$ $K\pi$ interactions, even though only the former is dominated by resonances. Here we estimate how big the fraction of $I=3/2$ amplitude is in $D$ decay.
\begin{figure}[bh]
\vspace{3mm}
\begin{center}
\includegraphics[width=15.cm]{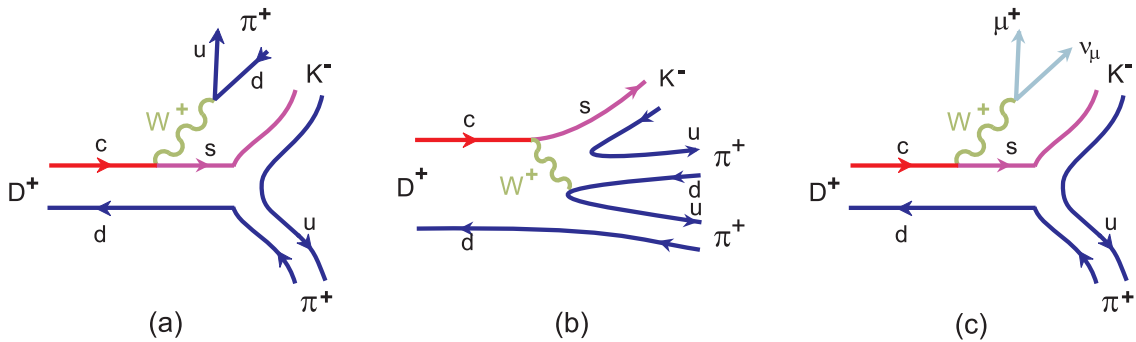}
\end{center}
 \caption{\small{Quark line diagrams for $D$-decays. (a) is the typical spectator model that underlies the idea that the resulting $K\pi$ system has just $I=1/2$. (b) involves internal $W$-boson conversion. For semi-leptonic decays only the analogue of (a) is possible. This is shown as (c).}}
\end{figure}

\noindent This is made possible by a recent model-independent treatment of the E791 results on $D\to K\pi\pi$ decay presented at La Thuile by Meadows~\cite{meadows}. There a Dalitz plot analysis is performed that isolates the phase and magnitude of the $K\pi$ $S$-wave interaction bin by bin in $K\pi$ mass. To explain how this allows us to estimate the $I=1/2$ and $3/2$ components, we begin with $K\pi\to K\pi$ scattering. For this the $S$-wave amplitude, in particular, is given by
\be
{\cal T}(K\pi\to K\pi;s)\;=\;C_{1/2}\,{\cal T}^{1/2}(s)\;+\;C_{3/2}\,{\cal T}^{3/2}(s)\quad , 
\ee
where $s$ is the square of the $K\pi$ mass, and the $C_I$ are the appropriate Clebsch-Gordan coefficients depending on the charges of the $K\pi$ system.  
The hadronic amplitude ${\cal T}^I$ with definite isospin has a magnitude and phase related by elastic unitarity for $\sqrt{s}\,<\,(m_K+m_{\eta'})$, the first strongly coupled threshold, and given by:
\be
{\cal T}^I(s)\;=\;\frac{1}{\rho}\,\sin \delta^I\, \exp(i \delta^I)\quad ,
\ee
where $\rho\,=\,2 k/\sqrt{s}$ with $k$ the $K\pi$ c.m. 3-momentum.
For the famous LASS experiment~\cite{lass} the $K^-\pi^+\to K^-\pi^+$ amplitude is thus given by
\be
{\cal T}(K^-\pi^+\to K^-\pi^+;s)\;=\;\frac{2}{3\rho}\,\left[\sin \delta^{1/2} \exp(i\delta^{1/2})\;+\frac{1}{2}\,\sin \delta^{3/2} \exp(i\delta^{3/2})\right]\quad .
\ee
These data, plus results on $K^-\pi^-\to K^-\pi^-$~\cite{estabrooks}, fix $\delta^I$ for both $I=1/2,\ 3/2$. The LASS group provide parametrisations~\cite{lass2} of these data, which are displayed as the dashed lines in the lower part of Fig.~2.
\begin{figure}[p]
\begin{center}
\includegraphics[width=10.cm]{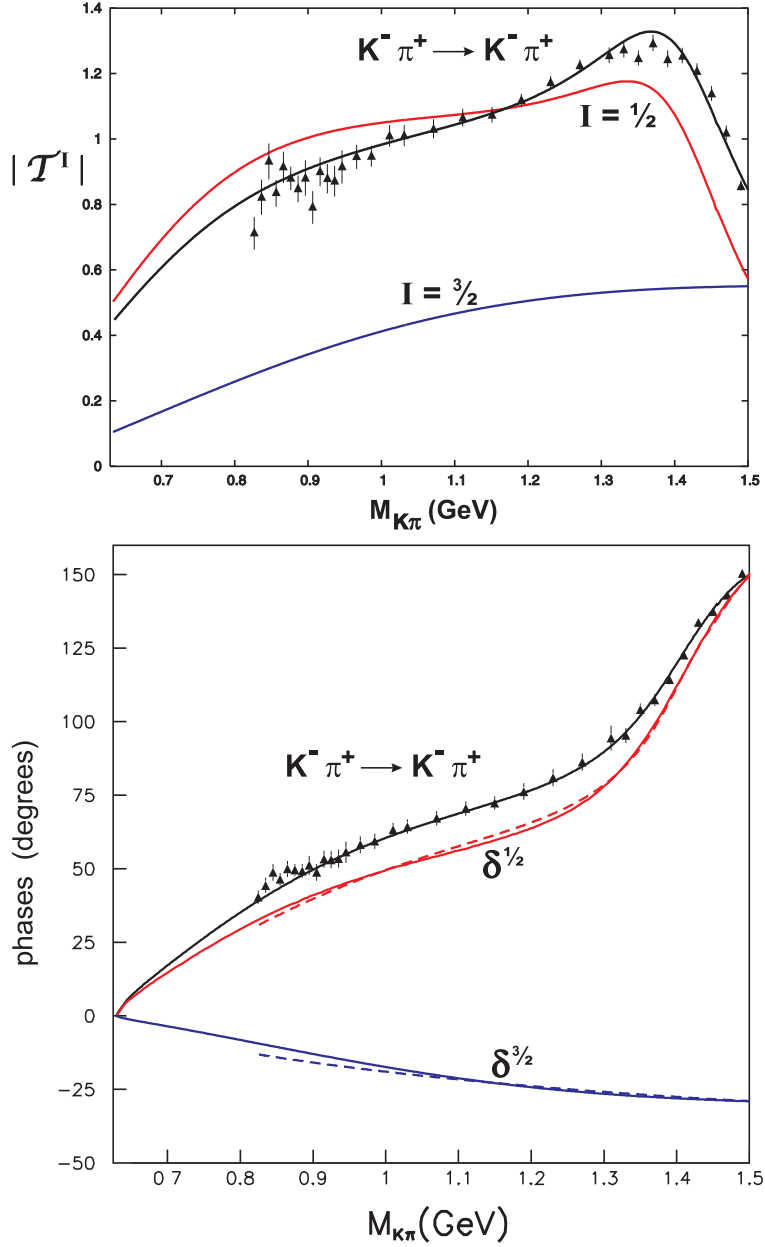}
\end{center}
 \caption{\small{In the upper and lower figures are the magnitudes and phases, respectively, of the $I=1/2$ and $3/2$ amplitudes, ${\cal T}^I$, given by fits (black line) to the LASS results on $K^-\pi^+\to K^-\pi^+$ scattering, shown as the data points. The
fits shown are built from the $I=1/2$ and $3/2$ phases $\delta^I$ according to Eq.~(3). The dashed lines are the LASS group's own parametrisation~\cite{lass2}. The solid lines~\cite{BP} display a representation that not only describes the data, but extrapolates to threshold in keeping with one loop Chiral Perturbation Theory~\cite{bernard}.}}
\end{figure}

\noindent Though these are valid in the experimental region LASS accessed, i.e.  above 825 MeV, their extrapolations down to $K\pi$ threshold are known to be inconsistent with Chiral Perturbation Theory.
Consequently, we consider a parametrisation~\cite{BP} of the elastic phases, shown as the solid lines in Fig.~2, that is consistent with the calculations of Chiral Perturbation Theory~\cite{bernard} as shown in the dispersive analysis by B\"uttiker {\it et al}.~\cite{descotes} These phases are the key ingredients in our discussion.

\noindent We now turn to $D$ decay, firstly in its semi-leptonic mode to $K\pi\mu\nu$. The $S$-wave amplitude in the $K\pi$ channel  is given by
\be
{\cal F}(D\to (K\pi)\mu\nu;s)\;=\;{\cal F}^{1/2}_{sl}(s)\;+\;{\cal F}^{3/2}_{sl}(s)\quad ,
\ee
where the ${\cal F}^I_{sl}(s)$ are the semileptonic production amplitudes with $I=1/2, 3/2$. In this case the emitted $W^+$ is the source of the dilepton system, which of course, has no strong interaction with the $K\pi$ pair, as illustrated in Fig.~1c.   
Now elastic unitarity requires that the phase of the $K\pi$ interaction with definite spin and isospin is the same as that in elastic $K\pi$ scattering, so that: 
\be
{\cal F}^{I}_{sl}(s)\;=\;\mid{\cal F}^{I}_{sl}(s)\mid\,\exp\left[i \delta^I(s)\right]\quad .
\ee
This is the famous final state interaction theorem of Watson~\cite{watson}.
Results from FOCUS~\cite{focus} confirm that this relationship holds and that the $I=3/2$ component is small or negligible in this process.

\noindent With this confirmation of Watson's theorem in heavy flavour decays, we consider the wholly hadronic channel: 
$D^+$ decay to $K^-\pi^+\pi^+$. 
This has been analysed by E791 in a \lq\lq new approach to the analysis of 3 body decays''~\cite{meadows}. 
The structure of the Dalitz plot has marked bands in both $K^-\pi^+$ mass combinations for the well known 
$K^*(892)$ and $K_2^*(1430)$ that feature so prominently in $K^-\pi^+\to K^-\pi^+$ scattering~\cite{estabrooks,lass}.
 This suggests the dominance of 2-body strong interactions. Consequently, E791 assume a generalised isobar 
picture in which genuine 3-body interactions are neglected. As is by now standard~\cite{cleo-c,focus-kmatrix}, the decay matrix element 
is represented in the way summarised, for instance, by the CLEO-c group in Section II of Ref.~\cite{cleo-c} and so for the $D$-decay  we consider here  is then a sum of the amplitudes for the two possible combinations of 
$K^-\pi^+$ interactions with their appropriate spin and vertex factors. As in the usual isobar picture, all waves with $J>0$
 are represented by a sum of resonance terms. This incorporates the known $I=1/2$ $K\pi$ resonances like the $K^*(892)$
 $K^*_1(1410)$ and $K^*_1(1680)$ for $J=1$, for instance. With these resonances being well separated, their amplitude can be reasonably 
described by a sum of Breit-Wigner forms. Each partial wave has a production phase from the $D\pi\to K^*$ coupling, which 
is assumed to be a constant as is also standard~\cite{cleo-c,meadows}. Since the parameters of each resonance are those 
found in $K\pi$ scattering, {\it e.g.} by LASS~\cite{lass}, the phase variation of each $K\pi$ partial wave in $D$-decay  
is exactly that of $K\pi$ scattering in the elastic region, in agreement with the application of Watson's theorem to such an isobar picture~\cite{aitchison}. 
Since one overall phase is not determinable, the strong $P$-wave is taken as the reference wave with its production phase 
set to zero. 
\begin{figure}[p]
\vspace{3mm}
\begin{center}
\includegraphics[width=12.cm]{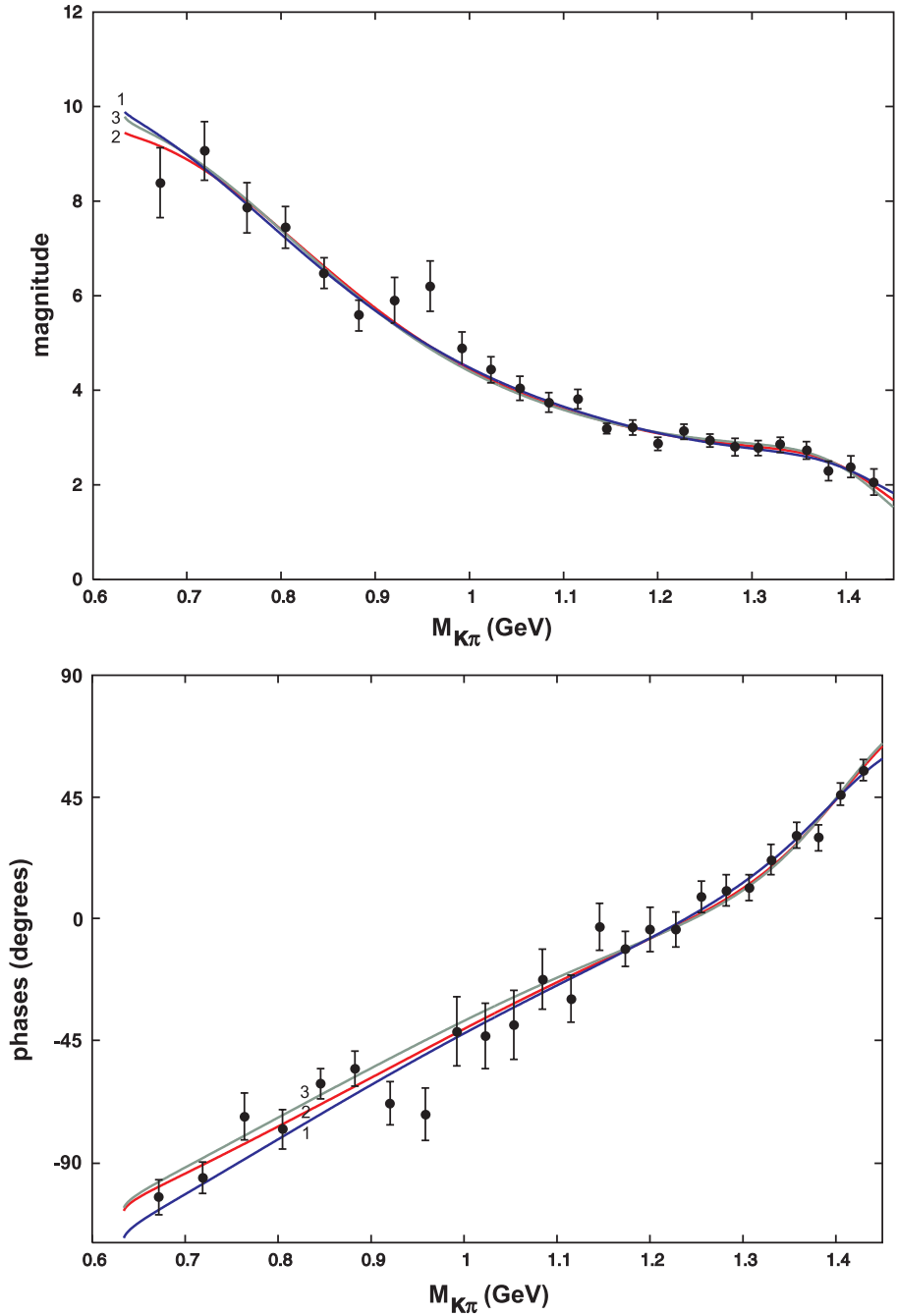}
\end{center}
\vspace{3mm}
\caption{\small{The data are the magnitude and phase of the $S$-wave amplitude determined by the E791 analysis of their $D$-decay results. The ``best'' fit to these data with a quadratic form for the $\alpha_I$ of Eq.~(18) is the {\it red} curve labelled 2, also shown in Fig.~4. Curves 1 and 3 in {\it blue} and {\it green} respectively mark one standard deviation away from the optimal fit. }}
\end{figure}

\noindent What is new in the E791 analysis is the description of the $K\pi$ $S$-wave. There the well-established wide 
$K^*_0(1430)$ appears, with perhaps an even broader $\kappa$ at lower mass. These are not simply describable by a sum of 
isolated Breit-Wigners, as the $K^-\pi^+\to K^-\pi^+$ amplitude of Fig.~2 illustrates. Rather than enforce some {\it ad hoc} 
prescription of this key wave, E791 represent this by a magnitude and phase in each bin of $K\pi$ mass in each $K^-\pi^+$ 
combination. The beauty of this analysis is that each $S$-wave $K\pi$ mass band overlaps with a crossed $K\pi$ band in a 
$P$-wave and so the relative phase is determined, as well as the magnitude of the $S$-wave. This provides as close to a 
model independent determination of the $K^-\pi^+$ $S$-wave interaction in this $D$-decay as is presently possible. 
The results are shown in Fig.~3. From the application of Watson's theorem to this generalised isobar picture~\cite{aitchison}, one 
would expect the phase variation of this $S$-wave amplitude to follow that of $K\pi$ scattering in the region of 
elastic unitarity. As already mentioned (just before Eq.~(2)), elastic unitarity is found~\cite{estabrooks,lass} to be very nearly exact upto $K\eta'$ threshold, despite 
the fact that this is above the opening of the $K\pi\pi$ and $K\eta$ channels, from which we infer these to be 
negligible below 1450 MeV. While the phase of the $K\pi$ $S$-wave in $D$-decay and that of $K\pi$ scattering (compare the 
lower graphs of Figs.~2 and 3) both have an upward trend in the \lq\lq elastic'' region, they do not match 
to the precision expected. The reasons for this can be  manifold:
\begin{itemize}
\item{} the isobar assumption of only 2-body $K\pi$ interactions may not be true,
\item{} even if it is, then perhaps Watson's theorem does not apply and so the phase variation is not required  
to be the same,
\item{} even if Watson's theorem applies, should the phase variation of the $K^-\pi^+$ interactions be that of 
$K^-\pi^+\to K^-\pi^+$ scattering or that of just its $I=1/2$ component?
\end{itemize}
These are all questions that have been raised at the BaBar Dalitz Workshop in December 2004~\cite{dalitz}.

\newpage
\baselineskip=5.45mm

\noindent Since the generalised isobar picture of pairwise $AB$ interactions defines a model for $AB \to AB$ scattering, 
we adopt the view that this must be in keeping with data on $AB\to AB$ scattering and the phase variation must agree in 
the region of elastic unitarity~\cite{aitchison}. This has to be so, for the model to be consistent. Thus the $K\pi$ $S$-wave phase 
variation found by E791 must agree with that for $K\pi\to K\pi$ scattering. However, unitarity is only diagonalised
by partial wave amplitudes with definite quantum numbers such as isospin, and not in general for individual charged channels, like $K^-\pi^+$.
 Consequently, it is the phase variation with definite isospin that should match. There is no reason that the 
combination of $I=1/2$ and $I=3/2$ contributions determined by simple Clebsch-Gordan coefficients in $K^-\pi^+$ scattering (as in Eq.~(3)) 
is that formed in $D$-decay, to which we already alluded in the introduction. 
The relationship between the phase variations then allows us to estimate the relative $I=1/2$ and $3/2$ components of the 
E791 $S$-wave, as we shall shortly describe. Of course, there could also be $I=3/2$ components in the higher 
partial waves too. These have been neglected in the E791 resonance-dominated description of these amplitudes. 
Nevertheless, it is known~\cite{estabrooks} that the $J \ge 1$ waves with $I=3/2$ vary by less than $3^o$ between $K\pi$
threshold and 1.8 GeV. Consequently, any such components should have only a tiny effect on our estimate.

\noindent In keeping with the generalised isobar description that E791 adopt, 
the effect of the spectator pion is to produce an additional production phase, $\beta_I$, which we take to be a constant, 
as they do for all other waves. The $K^-\pi^+$ $S$-wave determined by E791 we call ${\cal F}$, which is a function of
$s$, the square of the $K\pi$ invariant mass, $E$. This is then given by
\be
{\cal F}(s)\;=\;{\cal F}^{1/2}_{had}(s)\;+\;{\cal F}^{3/2}_{had}(s)\;\equiv\; {\cal A}(s)\,\exp[i\phi(s)]\quad ,
\ee
where the amplitudes with definite quantum numbers are given by:
\be
{\cal F}^{I}(s)_{had}\;=\;\mid{\cal F}^{I}_{had}(s)\mid\,\exp\left[i \delta^I(s)+ i \beta_I\right]\quad , 
\ee
in the region of elastic unitarity below 1450 MeV.
The phases $\beta_I$ reflect the structure of the complete set of quark line graphs of Figs.~1a,~b in the strong coupling limit 
before final state interactions are included. Since different graphs contribute to each $K\pi$ isospin, the production phase 
$\beta_I$ depends on $I$. It is the magnitude ${\cal A}$ and phase, $\phi$, of the total $S$-wave $K\pi$ amplitude, 
displayed in Fig.~3, that E791 have determined~\cite{meadows}.
Given this and the $K\pi$ phases, $\delta^I$ shown in Fig.~2, the aim is to deduce the magnitude of the $I=1/2$ and
 $3/2$ $K\pi$ amplitudes in $D$ decay.

\noindent Let us see how to do this. In vector terms it is like finding the vectors
${\bf {\cal F}^{1/2}}$ and ${\bf {\cal F}^{3/2}}$ knowing only their sum ${\bf {\cal A}}$, i.e.
\be
{\bf {\cal A}}\;=\;{\bf {\cal F}^{1/2}}\;+\;{\bf {\cal F}^{3/2}}\qquad .
\ee
Clearly there are an infinite number of vectors ${\bf {\cal F}^I}$ that satisfies this. If the production phases, $\beta_I$, were zero (or otherwise determined), the solution would be simple, since we know from $K\pi$ elastic scattering what the phases $\delta_I$ are at each energy in the elastic region, Fig.~2. Knowing the directions of the vectors, their magnitudes can easily be found. However, here we do not know in advance what the production phases are, but we do know that these are the same phases at every $K\pi$ mass, $E\,=\,\sqrt{s}$. This sets the scene for determining the amplitudes.

\noindent Starting from 
\be
{\cal A}\,\exp(i\phi)\;=\;{\cal F}^{1/2}\;+\;{\cal F}^{3/2} \qquad ,
\ee
where the ${\cal F}^I$ are the complex $I=1/2,\,3/2$ amplitudes,
it is straightforward to check that the solution is :
\begin{eqnarray}
{\cal F}^{1/2}(E)&=& {\cal A}\ \frac{\sin\left(\delta^{3/2}(E)+\beta_{3/2}-\phi(E)\right)}
{\sin\left(\delta^{3/2}(E)-\delta^{1/2}(E)-\beta_{1/2}+\beta_{3/2}\right)}
\;\exp\left[i(\delta^{1/2}(E)+\beta_{1/2})\right]\quad ,\\
\nonumber&&\\
{\cal F}^{3/2}(E)&=& {\cal A}\ \frac{\sin\left(\delta^{1/2}(E)+\beta_{1/2}-\phi(E)\right)}
{\sin\left(\delta^{1/2}(E)-\delta^{3/2}(E)+\beta_{1/2}-\beta_{3/2}\right)}
\;\exp\left[i(\delta^{3/2}(E)+\beta_{3/2})\right]\quad .
\end{eqnarray}
The phase difference, $\delta^{1/2}(E)-\delta^{3/2}(E)$, varies from 0 to $\sim 180^o$ in the elastic region below $K\eta'$ threshold, and so we see that the denominator in each expression, namely
\be
\sin\left(\delta^{1/2}(E)-\delta^{3/2}(E)+\beta_{1/2}-\beta_{3/2}\right)
\ee
will inevitably vanish at at least one energy in this region. Let this reference energy be $E_r$. Since the amplitudes ${\cal F}^{I}$ are finite at real energies,
the numerator must vanish at this same reference energy. This fixes the production angles. Thus
\begin{eqnarray}
\beta_{1/2}&=&\phi(E_r)\,-\,\delta^{1/2}(E_r)\,+\,m\pi\\
\beta_{3/2}&=&\phi(E_r)\,-\,\delta^{3/2}(E_r)\,+\,n\pi\quad ,
\end{eqnarray}
where $\,m,\, n\,$ are integers (including zero). Consequently, we can write the individual isospin amplitudes in terms of just one unknown parameter, the reference energy $E_r$, as:
\begin{eqnarray}
\nonumber
{\cal F}^{1/2}(E)&=& {\cal A}\ \frac{\sin\left(\phi(E)-\phi(E_r)-\delta^{3/2}(E)+\delta^{3/2}(E_r)\right)}
{\sin\left(\delta^{1/2}(E)-\delta^{3/2}(E)-\delta^{1/2}(E_r)+\delta^{3/2}(E_r)\right)}\\
&& \times
\exp\left[i(\delta^{1/2}(E)-\delta^{1/2}(E_r)+\phi(E_r))\right]\quad \\
\nonumber &&\\
\nonumber
{\cal F}^{3/2}(E)&=& {\cal A}\ \frac{\sin\left(\delta^{1/2}(E)-\delta^{1/2}(E_r)-\phi(E)+\phi(E_r)\right)}
{\sin\left(\delta^{1/2}(E)-\delta^{3/2}(E)-\delta^{1/2}(E_r)+\delta^{3/2}(E_r)\right)}\\
&&\times \exp\left[i(\delta^{3/2}(E)-\delta^{3/2}(E_r)+\phi(E_r))\right]\quad .
\end{eqnarray}
\begin{figure}[p]
\vspace{3mm}
\begin{center}
\includegraphics[width=12.cm]{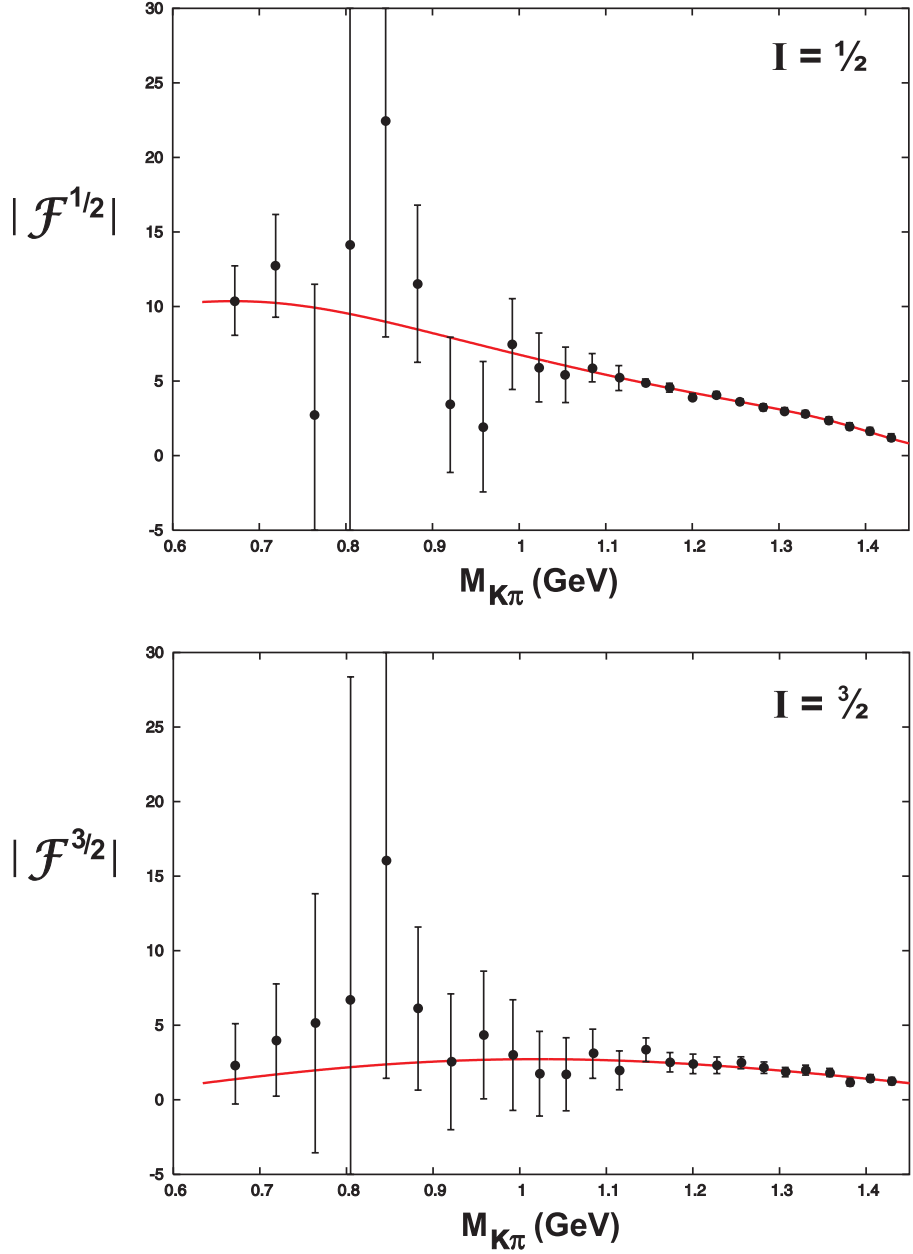}
\end{center}
 \caption{\small{The moduli of the $I=1/2$ and $3/2$ $K\pi$ components of the $D$-decay amplitude determined from the E791 analysis using Eqs.~(15,16) with $E_r\simeq 800$ MeV. The error bars shown are fixed by the uncertainties given by E791. They are inevitably largest in the region of $E\,\simeq\, E_r$. The fit given by representing the coupling functions $\alpha_I$ as quadratics in $s$ is shown. }}
\end{figure}
For each value of $E_r$ we can then determine the amplitudes. Consequently, we have a continuous range of possible amplitudes. Within this set, there are only a small group that are physically (as opposed to mathematically) allowed.
Watson's theorem is ensured by the phase variation of each of ${\cal F}^{1/2}$ and ${\cal F}^{3/2}$ being given by the elastic phases of the corresponding elastic scattering amplitudes ${\cal T}^I$. However, the dynamics of the final state interactions requires that the amplitudes ${\cal F}^I$ should be smoothly connected to the scattering amplitudes ${\cal T}^I$ through coupling functions ${\overline{\alpha_I}}(E)$. A consequence of this is that any resonance poles in $K\pi$ scattering appear in the decay process with the same mass and width. The functions ${\overline{\alpha_I}}$ fix the coupling to the decay channel.
Thus
\be
{\cal F}^I(E)\;=\;{\overline{\alpha_I}}(E)\ {\cal T}^I(E) ~\exp(i\beta_I)\quad .
\ee
The dynamics of the pseudoscalar interactions imposes Adler zeros in each
of the $K\pi$ scattering amplitudes with $\,I\,=\,1/2,\,3/2\,$ at $\,s\,=\,s_I^{\, 0}\,$. Such zeros may appear in the decay process but not necessarily at the same position as in  elastic scattering. We take account of this by defining reduced elastic scattering amplitudes with the Adler zero divided out. We then specify
new coupling functions, $\,\alpha_I\,$ which have to be smooth  since they contain no explicit $s-$channel dynamics. Thus we have
\be
{\cal F}^I(E)\;=\;\alpha_I(E)\ \frac{{\cal T}^I(E)}{s-s_I^{\,0}}~\exp(i\beta_I) \quad ,
\ee
where the coupling functions $\,\alpha_I\,$ are representable by low order polynomials in $\,s\,=\,E^2\,$. 

\noindent
We therefore determine the coupling functions $\alpha_I(E)$ for each choice of $E_r$ and select those that are representable by low order polynomials, while still providing an accurate fit to the amplitude determined by the E791 analysis. 
With 26 values of magnitude and  26 phases in the elastic region below $K\eta'$ threshold, this is a severe constraint.

\noindent The \lq\lq best'' fit is found with $E_r\,\simeq\,800$ MeV, for which
the production phases are $\beta_{1/2}\,=\,72^o$ and $\beta_{3/2}\,=\,-73^o$. 
The moduli of the
$I=1/2$ and $3/2$ amplitudes, using Eqs. (15,16), are shown in Fig.~4.
As can be seen these are accurately determined away from $E_r$, where the errors are very small. These show that the $\,\alpha_I\,$ of Eq.~(18) are well represented by quadratic polynomials in the square of the $K\pi$ mass, $s$. The precise results above 1 GeV fix the continuation to lower $K\pi$ mass as illustrated in Fig.~3. The requirement that the $D$-decay amplitude of Eq.~(6) should be well described is illustrated by
the (red) lines, labelled 2 in Fig.~3. Allowing a one standard deviation change  by varying $E_r$ alters the $I=1/2$ and $3/2$ amplitudes, while still fitting the E791 magnitudes and phase. This result is shown in Figs.~3,~5. The lines 1,~3 (in blue and green, respectively) in Fig.~3 show the corresponding small difference in
the description of the E791 amplitude, while Fig.~5 shows the range of variation of the $I=1/2$ and $3/2$ then permitted.
The bands reflect a change of $E_r$ from 750 to 950 MeV, or equivalently a range of
\be
\beta_{1/2}\,=\,65^o\,\to\, 85^o \quad,\quad\beta_{3/2}\,=\, -86^o\,\to\,-38^o\qquad ,
\ee
which are strongly correlated.

\begin{figure}[h]
\begin{center}
\includegraphics[width=12.cm]{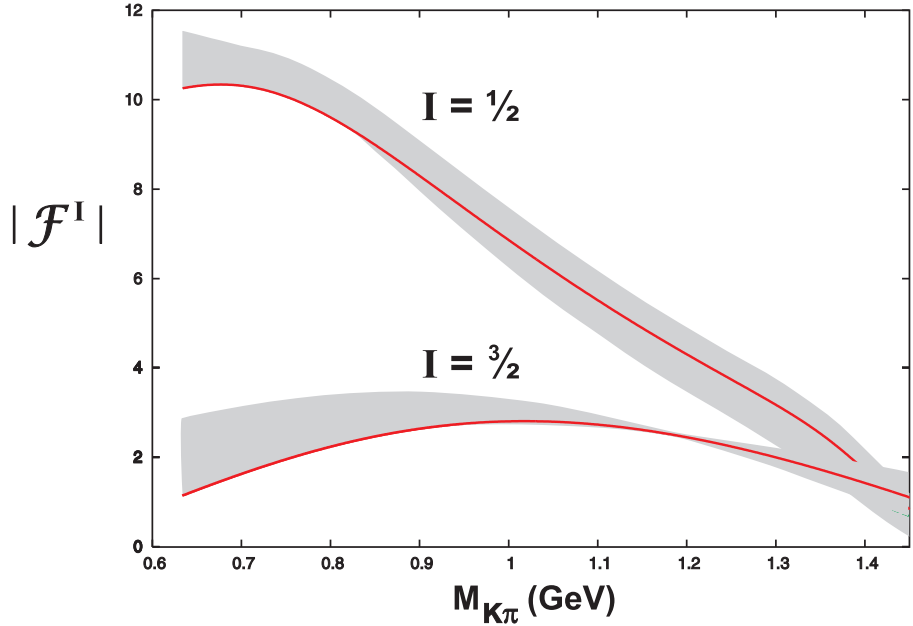}
\vspace{-3mm}
\end{center}
 \caption{\small{The {\it red } (solid) lines give the magnitudes of the $I=1/2$ and $3/2$ components of the $D$-decay amplitude determined by the \lq\lq best'' fit to the data of Fig.~3. The bands delineate the ranges set by one standard deviation variation from this \lq\lq best'' fit.   }}
\end{figure}

\noindent We, of course, only know that the coupling functions $\alpha_I$ of Eq.~(18) should be smooth functions of $K\pi$ mass since they do not contain direct
$K\pi$ dynamics. A constant or linear function of $s$ does not give an acceptable fit in terms of $\chi^2$ for any value of $E_r$. A quadratic is the lowest order polynomial to give fits of acceptable confidence, as seen from the curvature of the magnitudes in Fig.~4 determined by the data-points with small error bars above 1.1 GeV, as shown in Fig.~3.
 Adding higher order terms in the polynomial representation of the $\alpha_I$ does not significantly improve the confidence level.

\noindent We see by comparing the upper plot of Fig.~2 with Fig.~5 that the $I=1/2$ and $3/2$ $K\pi$ components of $D$ decay
are quite different from those of elastic scattering.  The $I=3/2$ component is more than 50\% of the $I=1/2$ above 1.1 GeV. This is surprising because the
$I=1/2$ amplitude contains the broad $K^*_0(1430)$, 
while the $I=3/2$ is entirely non-resonant. Consequently, any analysis of the $D$-decay data which neglects the $I=3/2$ amplitude must place this component
incorrectly in other contributions and so lead to  false conclusions about resonant
branching fractions. Neither isospin component appears to contain an Adler zero and hence they grow towards threshold. At low mass we see the $I=1/2$ amplitude is dominant, whether this is because there is a near threshold $\kappa$ resonance requires a far better determination of the amplitude and phases.                     

\noindent An important ingredient in  fixing the isospin components of the $D$-decay amplitude are the phases of $K\pi$ scattering in the elastic region. Here these
have been set by the LASS experiment above 825 MeV and their continuation down to threshold by the predictions of Chiral Perturbation Theory.
However, increased statistics on $D\to K\pi\pi$ decay would reduce the error bars on the amplitude and phases in Fig.~3. These might well become sufficiently small that an acceptable fit cannot be obtained without changing the $K\pi$ phases $\delta^I$ away from their LASS results. This would mean that the $D$-decay results could indeed increase the precision within which the $K\pi$ elastic phases are known. Such improvement in $D$-decay statistics would be particularly welcome below 1 GeV down to threshold, where the current errors are sizeable --- Figs.~4,~3. There it holds out the prospect of revealing whether there is indeed a 
low mass $\kappa$ or not. Such a state can only be exposed by analytic continuation of the appropriate amplitudes into the complex $s$-plane to deduce whether a pole exists or not. This requires far greater precision than the E791 data and its analysis we have considered here. $D$-decay results from $B$-factories may make this possible. The present analysis points the way. 

\newpage

\noindent {\bf Acknowledgements}

\noindent We are most grateful to Brian Meadows for discussions that prompted this analysis.
We acknowledge the key support of the EU-RTN Programme, 
Contract No. HPRN-CT-2002-00311, \lq\lq EURIDICE''. One of us (MRP) is grateful to the hospitality provided by the LNF 
Spring Institute Programme when this work was completed.

\end{document}